%% file: main.tex
\documentclass[journal]{IEEEtran}

\usepackage{graphicx}
\usepackage{amsmath,amssymb}
\usepackage{booktabs}
\usepackage{siunitx}
\usepackage{array}
\usepackage{tabularx}
\usepackage{cite}
\usepackage{hyperref}
\usepackage{cleveref}
\usepackage{tikz}
\usetikzlibrary{arrows.meta,positioning,shapes.geometric}

\crefname{figure}{Fig.}{Figs.}
\Crefname{figure}{Fig.}{Figs.}
\crefname{table}{Table}{Tables}
\Crefname{table}{Table}{Tables}
\crefname{equation}{Eq.}{Eqs.}
\Crefname{equation}{Eq.}{Eqs.}

\newcolumntype{Y}{>{\raggedright\arraybackslash}X}
\newcolumntype{L}[1]{>{\raggedright\arraybackslash}p{#1}}

\hypersetup{
  colorlinks=true,
  linkcolor=blue,
  citecolor=blue,
  urlcolor=blue
}

\graphicspath{{analysis/}}

\title{Reliability-Aware Bayesian Optimization of 1310-nm PCSELs with FDTD Verification}

\author{Jinglin~Yu,
        Feiyang~Wu,
        Longying~Wen,
        Chongxian~Yuan,
        Renjie~Li,
        and Zhaoyu~Zhang%
\thanks{This research was supported by the Shenzhen Science and Technology Program under Grant No. KJZD20230923115114027, the Guangdong Key Laboratory of Optoelectronic Materials and Chips under Grant No. 2022KSYS014, the Shenzhen Key Laboratory Project under Grant No. ZDSYS201603311644527, the Longgang Key Laboratory Project under Grant Nos. ZSYS2017003 and LGKCZSYS2018000015, and the Innovation Program for Quantum Science and Technology (Grant No. 2021ZD0300701), Hefei National Laboratory, Hefei 230088, China.}
\thanks{Jinglin Yu, Feiyang Wu, Longying Wen, Chongxian Yuan, and Zhaoyu Zhang are with the School of Science and Engineering (SSE), The Chinese University of Hong Kong, Shenzhen, Shenzhen 518172, China.}
\thanks{Renjie Li is with the Nick Holonyak Micro and Nanotechnology Laboratory, University of Illinois Urbana-Champaign, Urbana, IL 61801 USA.}
\thanks{Jinglin Yu, Feiyang Wu, and Longying Wen contributed equally to this work. Corresponding authors: Renjie Li (renjie2@illinois.edu) and Zhaoyu Zhang (zhangzy@cuhk.edu.cn). E-mail: 121090729@link.cuhk.edu.cn; fywu2003@gmail.com; 225015041@link.cuhk.edu.cn; cxyuan@cuhk.edu.cn.}}

\begin{document}

\maketitle

\begin{abstract}
Near-\SI{1310}{nm} photonic-crystal surface-emitting lasers (PCSELs) are attractive narrow-beam sources for optical communication and sensing, but their final design refinement is costly. Small geometry changes simultaneously shift the band-edge resonance, cavity leakage, far-field divergence, and the numerical stability of a high-$Q$ decay fit, while every full-wave trial requires a time-domain simulation. We couple a commercial finite-difference time-domain solver to a reliability-aware Bayesian optimization (BO) loop over eight local design variables. Each completed simulation updates the surrogate used to choose the next geometry. Candidate ranking combines wavelength and beam-quality requirements with a reliability-adjusted metric $Q_{\mathrm{eff}}$ derived from the solver-reported relative fit-error estimate $dQ/Q$. Across three 80-evaluation runs from the same reference model, BO produced 5--15 candidates per run that passed the joint filter. Designs reconstructed from fresh model copies retained $Q_{\mathrm{eff}}=4.33\times10^6$--$7.76\times10^6$, a 60--108-fold increase over the baseline metric, at 1308.23--1310.90~nm with approximately $0.84^{\circ}$ divergence. Under equal budgets, BO gave the highest mean strict-filter yield (9.0 candidates), compared with differential evolution (7.0) and Latin-hypercube sampling (1.5), although the controls occasionally matched the peak $Q_{\mathrm{eff}}$. Field maps, resonance spectra, and local perturbations further identify an index-related wavelength handle and a hole-size-related leakage handle. The resulting FDTD budget produces a pool of wavelength-compatible, narrow-beam, and reproducible high-$Q$ PCSEL candidates without trusting a single optimistic decay fit.
\end{abstract}


\input{sections/01_introduction}
\input{sections/02_device_and_problem}
\input{sections/03_methods_bo}
\input{sections/04_results_bo}
\input{sections/05_conclusion_bo}
\input{sections/06_appendix}
\input{sections/07_declarations}

\bibliographystyle{IEEEtran}
\bibliography{references}

\end{document}

%% file: sections/01_introduction.tex
\section{Introduction}

Photonic-crystal surface-emitting lasers (PCSELs) use two-dimensional band-edge feedback to combine broad-area coherent oscillation with surface-normal emission \cite{Imada1999CoherentPCSEL,Noda2001PolarizationPCSEL,Noda2017JSTQEPCSEL,Sakurai2019ProgressPCSEL,Noda2023TutorialPCSEL}. The scalable emitting aperture, narrow divergence, and control of the radiated beam have enabled watt-class sources, beam steering, and LiDAR-oriented devices \cite{Hirose2014WattClassPCSEL,Yoshida2019DoubleLatticePCSEL,Sakata2020DuallyModulatedPCSEL,Yoshida2021LiDARPCSEL,Inoue2022GeneralRecipePCSEL}. Short-pulse and small-signal modulation have also been demonstrated \cite{Morita2021GainLossPCSEL,Orchard2023SmallSignalPCSEL}, while GaN and continuous-wave InP devices extend PCSEL operation from visible wavelengths to the \SI{1.3}{\micro\metre} communication band \cite{Matsubara2008GaNPCSEL,Itoh2020InPPCSEL}. These capabilities make the final adjustment of wavelength, cavity lifetime, and beam quality a device-design problem rather than a numerical afterthought.

The photonic-crystal pattern provides both in-plane feedback and out-of-plane radiation. A small change in layer thickness, effective index, lattice scale, or hole size can therefore move the band-edge resonance while changing vertical leakage and the far-field lobe. For an existing \SI{1310}{nm} template, exhaustive sweeps over these coupled variables are expensive because every trial requires a full finite-difference time-domain (FDTD) simulation and a decay-fit analysis. Bayesian optimization (BO) is suited to this setting: it uses the completed solver record to select the next geometry instead of fixing the entire sampling plan in advance \cite{Jones1998EGO,Snoek2012PracticalBO,Shahriari2016BayesianOptimization}.

Photonic inverse design has been approached with adjoint and topology optimization, learned surrogates, and automated cavity search \cite{Jensen2011TopologyNanophotonics,LalauKeraly2013AdjointShape,Piggott2015WDMInverseDesign,Molesky2018InverseDesign,Christiansen2021TopOptTutorial,Minkov2014AutomatedPhCCavities,Asano2019NanocavityDNN,Ma2021DeepLearningPhotonics,Liu2021MLPhotonicInverseDesign,Pan2023DeepAdjointReview,Sanchez2024MLOptimizationPhotonics}. Here the problem is narrower: eight scripted offsets modify an inherited commercial FDTD model inside a local design box. A second difficulty appears at high fitted $Q$. The finite-time decay analysis can report a large value together with a large fit-error estimate $dQ$, so maximizing raw $Q$ may favour a numerically unstable resonance rather than a dependable device candidate.

We address this local refinement problem with an FDTD-coupled, reliability-aware BO workflow. A relative fit-error penalty converts the fitted $Q$ into $Q_{\mathrm{eff}}$, while wavelength and far-field requirements remain explicit. Three repeated BO runs, two changed starting templates, and equal-budget controls measure search behaviour. Selected geometries are then rebuilt from untouched source-model copies and examined through near fields, far fields, resonance maps, solver-setting changes, and local perturbations. Only these reconstructed runs supply the reported device values. They yield a set of near-\SI{1310}{nm}, sub-degree, million-level-$Q_{\mathrm{eff}}$ PCSEL structures under the same full-wave post-processing used for the reference device.

%% file: sections/02_device_and_problem.tex
\section{PCSEL Structure and Optimization Problem}
\label{sec:device_problem}

\subsection{Device Structure and Design Variables}

All simulations begin with the recovered commercial FDTD file \texttt{PCSEL-1310-GaAs100\_origin.fsp}. The optimizer only edits copied files, so the source model remains unchanged during the search.

A single completed FDTD call includes a time-domain run, resonance selection, Q analysis, and far-field post-processing. The retained structures are judged in the \SI{1310}{nm} band by three quantities that move together in practice: resonance position, cavity quality, and emitted-beam width.

\begin{figure*}[!t]
  \centering
  \begin{tikzpicture}[
    font=\small,
    line join=round,
    >=Latex,
    arrow/.style={-{Latex[length=2.0mm]}, thick},
    feedback/.style={<->, thick, dashed, draw=blue!65!black}
  ]
    \begin{scope}[
      x={(0.82cm,0.20cm)},
      y={(-0.54cm,0.29cm)},
      z={(0cm,0.72cm)}
    ]
      \filldraw[fill=gray!18, draw=gray!65]
        (-2.7,-1.45,0.36) -- (2.7,-1.45,0.36) --
        (2.7,1.45,0.36) -- (-2.7,1.45,0.36) -- cycle;
      \filldraw[fill=gray!32, draw=gray!65]
        (-2.7,-1.45,0) -- (2.7,-1.45,0) --
        (2.7,-1.45,0.36) -- (-2.7,-1.45,0.36) -- cycle;
      \filldraw[fill=gray!42, draw=gray!65]
        (2.7,-1.45,0) -- (2.7,1.45,0) --
        (2.7,1.45,0.36) -- (2.7,-1.45,0.36) -- cycle;

      \filldraw[fill=teal!18, draw=teal!65!black]
        (-2.7,-1.45,0.58) -- (2.7,-1.45,0.58) --
        (2.7,1.45,0.58) -- (-2.7,1.45,0.58) -- cycle;
      \filldraw[fill=teal!38, draw=teal!65!black]
        (-2.7,-1.45,0.36) -- (2.7,-1.45,0.36) --
        (2.7,-1.45,0.58) -- (-2.7,-1.45,0.58) -- cycle;
      \filldraw[fill=teal!48, draw=teal!65!black]
        (2.7,-1.45,0.36) -- (2.7,1.45,0.36) --
        (2.7,1.45,0.58) -- (2.7,-1.45,0.58) -- cycle;

      \filldraw[fill=blue!14, draw=blue!65!black]
        (-2.7,-1.45,0.84) -- (2.7,-1.45,0.84) --
        (2.7,1.45,0.84) -- (-2.7,1.45,0.84) -- cycle;
      \filldraw[fill=blue!32, draw=blue!65!black]
        (-2.7,-1.45,0.58) -- (2.7,-1.45,0.58) --
        (2.7,-1.45,0.84) -- (-2.7,-1.45,0.84) -- cycle;
      \filldraw[fill=blue!42, draw=blue!65!black]
        (2.7,-1.45,0.58) -- (2.7,1.45,0.58) --
        (2.7,1.45,0.84) -- (2.7,-1.45,0.84) -- cycle;

      \foreach \xx in {-2.15,-1.43,...,2.15} {
        \foreach \yy in {-1.0,-0.5,...,1.0} {
          \filldraw[fill=white, draw=blue!70!black, line width=0.25pt]
            (\xx,\yy,0.855) circle[radius=0.105];
        }
      }

      \draw[feedback] (-2.25,0,1.03) -- (2.25,0,1.03);
      \draw[feedback] (0,-1.12,1.03) -- (0,1.12,1.03);
      \foreach \xx in {-0.75,0,0.75} {
        \draw[arrow, draw=orange!85!black] (\xx,0,0.96) -- (\xx,0,2.02);
      }
    \end{scope}
    \node[font=\bfseries] at (0,-1.20) {(a) Perspective device view};
  \end{tikzpicture}

  \vspace{0.25em}

  \begin{tikzpicture}[
    scale=0.92,
    transform shape,
    font=\small,
    >=Latex,
    layer/.style={draw, minimum width=7.6cm, minimum height=0.56cm, align=center},
    arrow/.style={-{Latex[length=2.2mm]}, thick},
    callout/.style={align=left, inner sep=2pt, anchor=west},
    var/.style={draw, rounded corners=2pt, align=center, inner sep=4pt, text width=2.85cm}
  ]
    \node[layer, fill=gray!18, draw=gray!65] (substrate) at (0,0) {};
    \node[layer, fill=teal!18, draw=teal!65!black] (active) at (0,0.62) {};
    \node[layer, fill=blue!14, draw=blue!65!black] (phc) at (0,1.24) {};

    \foreach \x in {-3.15,-2.45,-1.75,-1.05,-0.35,0.35,1.05,1.75,2.45,3.15} {
      \draw[fill=white, draw=blue!70!black] (\x,1.30) circle (0.095);
      \draw[fill=white, draw=blue!70!black] (\x+0.35,1.48) circle (0.075);
    }

    \node[callout, text=blue!65!black] at (4.05,1.24) {photonic-crystal layer};
    \node[callout, text=teal!65!black] at (4.05,0.62) {active / guiding region};
    \node[callout, text=gray!70!black] at (4.05,0.00) {substrate / lower cladding};

    \draw[arrow, draw=orange!85!black] (0,1.67) -- (0,2.65)
      node[above, text=orange!70!black] {surface emission};
    \draw[arrow, draw=blue!65!black] (-3.35,2.05) -- (-0.85,2.05)
      node[midway, above, text=blue!65!black] {lattice modulation};
    \draw[arrow, draw=blue!65!black] (3.35,2.05) -- (0.85,2.05);

    \node[var, fill=blue!7, draw=blue!60!black] (lateral) at (-3.35,-1.15)
      {$\texttt{netDLen}$, $\texttt{netDA}$\\lateral and period\\offsets};
    \node[var, fill=teal!8, draw=teal!60!black] (thick) at (0,-1.15)
      {$\texttt{netDT}$, $\texttt{netDT1}$, $\texttt{netDT3}$\\thickness offsets};
    \node[var, fill=orange!8, draw=orange!75!black] (index) at (3.35,-1.80)
      {$\texttt{netDN1}$, $\texttt{netDN3}$, $\texttt{netDLeng}$\\index and hole-size\\offsets};
    \draw[arrow, draw=blue!65!black] (lateral.north) -- (-2.5,0.05);
    \draw[arrow, draw=teal!65!black] (thick.north) -- (0,0.62);
    \draw[arrow, draw=orange!75!black] (index.north) -- (2.2,1.35);
    \node[font=\bfseries] at (0,-2.75) {(b) FDTD model and scripted offsets};
  \end{tikzpicture}
  \caption{PCSEL device and local optimization variables. (a) The perspective view shows the patterned photonic-crystal layer, two-dimensional in-plane feedback, and surface-normal radiation. (b) The eight scripted offsets act on the unit-cell geometry, layer thicknesses, and refractive indices of the source FDTD model. The drawings are schematic and not to scale.}
  \label{fig:pcsel_structure}
\end{figure*}
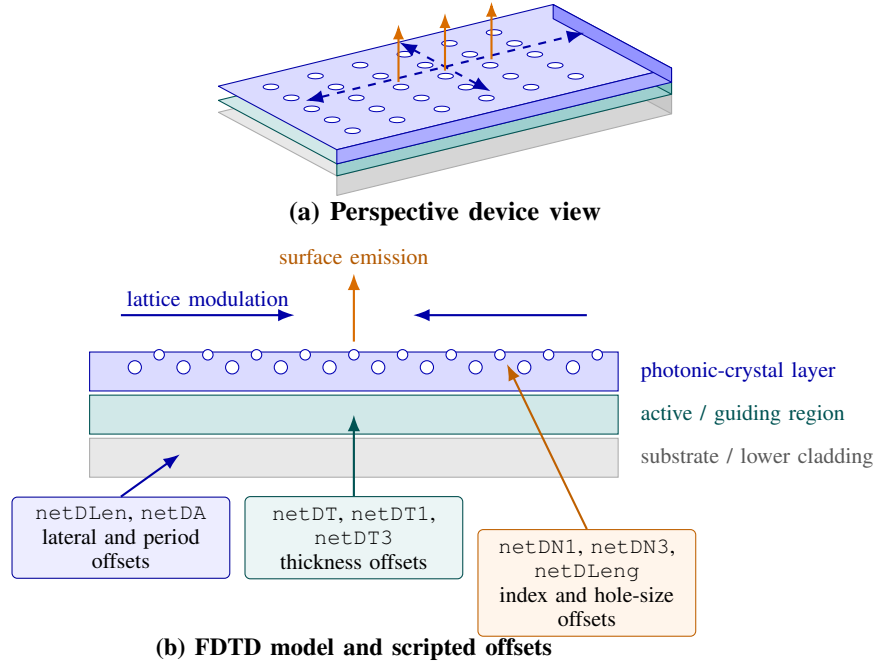

In the actual FDTD file, these handles are script variables inside the \texttt{pcsel} model object. The optimizer writes the offsets there and lets the source script regenerate the geometry.

The optimizer changes eight local offsets relative to the source template: \texttt{netDLen}, \texttt{netDT}, \texttt{netDT1}, \texttt{netDT3}, \texttt{netDN1}, \texttt{netDN3}, \texttt{netDLeng}, and \texttt{netDA}. \Cref{tab:design_variables} lists the search ranges used in the origin-template BO run.

\begin{table}[!t]
  \centering
  \footnotesize
  \caption{Design variables and bounds used for the origin-template BO run.}
  \label{tab:design_variables}
  \begin{tabularx}{\linewidth}{L{0.22\linewidth}Y L{0.19\linewidth}Y}
    \toprule
    Variable & Role in script & Origin BO range & Unit / interpretation \\
    \midrule
    \texttt{netDLen} & Length-related geometric offset & $[-130,130]$ & nm-scale offset \\
    \texttt{netDT} & Thickness-related offset & $[-35,35]$ & nm-scale offset \\
    \texttt{netDT1} & Layer/local thickness offset & $[-90,90]$ & nm-scale offset \\
    \texttt{netDT3} & Layer/local thickness offset & $[-60,60]$ & nm-scale offset \\
    \texttt{netDN1} & Refractive-index offset & $[-0.025,0.025]$ & dimensionless \\
    \texttt{netDN3} & Refractive-index offset & $[-0.035,0.035]$ & dimensionless \\
    \texttt{netDLeng} & Normalized hole-size offset & $[-2.4,2.4]$ & script-scaled offset \\
    \texttt{netDA} & Lattice/period offset & $[-15,15]$ & nm-scale offset \\
    \bottomrule
  \end{tabularx}
\end{table}

Length-like offsets use the scripted geometry units of the FDTD model; refractive-index offsets are dimensionless. The geometry script internally scales \texttt{netDLeng} before applying it to the hole-size factor. All BO results are perturbations of the recovered origin template.

\subsection{Parameter-to-Metric Coupling}

Near the band edge, the two-dimensional dielectric modulation couples counter-propagating in-plane waves and opens a channel for coherent vertical radiation. The relevant Fourier components of the unit cell therefore influence both feedback and out-coupling \cite{Noda2017JSTQEPCSEL,Inoue2020EnhancedFeedbackPCSEL,Inoue2022GeneralRecipePCSEL}. The resonance wavelength, cavity lifetime, and far-field pattern cannot be tuned independently in this structure.

The eight offsets act on different parts of this coupling. Thickness and refractive-index changes alter the modal optical path and confinement; within the present local box, increasing the index-related offset \texttt{netDN3} tends to red-shift the selected band-edge resonance. Lattice and hole-size changes alter the air-fill fraction and dielectric Fourier components. The scripted hole-size variable \texttt{netDLeng} consequently changes vertical leakage and the distribution of radiated power. Because the lateral aperture is unchanged, these unit-cell edits mainly adjust loss and beam formation rather than the device footprint.

The response is local and non-monotonic. Suppressing one radiation channel may raise the fitted $Q$, but the same perturbation can detune the resonance or move the Q-analysis routine to a neighbouring peak. This is why wavelength, far-field width, field pattern, and $dQ/Q$ are retained alongside $Q$. The parameter audit below identifies useful directions in this particular design box; it is not used as a substitute for a mode-resolved band-structure or coupled-wave calculation.

\begin{figure*}[!t]
  \centering
  \begin{tikzpicture}[
    node distance=1.0cm and 1.05cm,
    block/.style={
      rectangle,
      draw=blue!55!black,
      fill=blue!7,
      rounded corners=2pt,
      align=center,
      text width=3.15cm,
      minimum height=1.05cm,
      font=\small
    },
    metric/.style={
      rectangle,
      draw=orange!70!black,
      fill=orange!9,
      rounded corners=2pt,
      align=center,
      text width=2.65cm,
      minimum height=0.72cm,
      font=\small
    },
    arrow/.style={-{Latex[length=2.0mm]}, thick, draw=gray!65!black}
  ]
    \node[block, draw=teal!70!black, fill=teal!9] (offsets) {Scripted offsets\\thickness, index, lattice, hole size};
    \node[block, draw=violet!70!black, fill=violet!8, right=of offsets] (physics) {Optical response\\effective path, feedback,\\vertical leakage};
    \node[metric, draw=red!70!black, fill=red!8, right=of physics, yshift=1.25cm] (lambda) {resonance\\$\lambda$};
    \node[metric, draw=blue!65!black, fill=blue!8, right=of physics] (quality) {cavity quality\\$Q$, $dQ/Q$};
    \node[metric, draw=green!55!black, fill=green!9, right=of physics, yshift=-1.25cm] (beam) {far-field beam\\$\theta_{\mathrm{div}}$};
    \node[block, draw=orange!75!black, fill=orange!10, right=of quality] (selection) {Candidate selection\\high $Q_{\mathrm{eff}}$ near \SI{1310}{nm} with low divergence};

    \draw[arrow] (offsets) -- (physics);
    \draw[arrow] (physics.east) -- (lambda.west);
    \draw[arrow] (physics.east) -- (quality.west);
    \draw[arrow] (physics.east) -- (beam.west);
    \draw[arrow] (lambda.east) -- (selection.west);
    \draw[arrow] (quality.east) -- (selection.west);
    \draw[arrow] (beam.east) -- (selection.west);
  \end{tikzpicture}
  \caption{Color-coded physical map of the local PCSEL search. Scripted perturbations change wavelength placement, cavity quality, Q-fit reliability, and far-field divergence at the same time, so the retained candidate is selected by a coupled criterion instead of raw $Q$ alone.}
  \label{fig:physical_coupling}
\end{figure*}
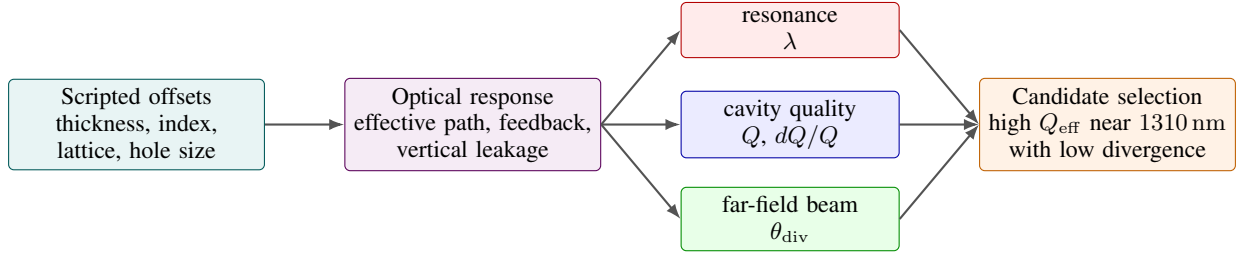

\subsection{Metrics and Optimization Target}

FDTD post-processing entails extracting the resonant wavelength $\lambda$, the fitted quality factor $Q$, Q-fit error estimate $dQ$, and the far-field divergence angle $\theta_{\mathrm{div}}$. In the Lumerical high-$Q$ analysis object, the value $Q$ is derived from the slope of the logarithmic time-domain decay after frequency isolation. The same analysis returns an error estimate for the fitted value \cite{AnsysQFactorCavity}. We treat $dQ/Q$ as a finite-time fit-reliability indicator. The online score rewards the wavelength alignment, low divergence, and fitted Q values that pass this error check. One score requires a full simulation, so later candidates should refer to the information from earlier runs.

\subsection{FDTD Evaluation and Verification}

After each simulation, the pipeline finds candidate resonances, extracts wavelength and Q-analysis data, and calculates the far-field divergence at the selected resonance. Direction-cosine axes are transformed into angular axes. The x- and y-direction intensity cuts are normalized, and the full-angle FWHM is obtained from half-maximum crossings by linear interpolation. The mean of the two FWHM values is used as $\theta_{\mathrm{div}}$.

The baseline PCSEL structure is evaluated first, and then Bayesian optimization is performed. After BO identifies candidate structures, the corresponding parameter sets are loaded into new FDTD copies. These standalone checks give the final wavelength, $Q$, $dQ$, and divergence-angle values used in the tables and discussion.

%% file: sections/03_methods_bo.tex
\section{FDTD-Coupled Bayesian Optimization Method}
\label{sec:methods}

\subsection{Closed-Loop FDTD Evaluation}

Given a proposed vector, the run script takes the recovered template and writes eight offsets into the \texttt{pcsel} model object, performs time-domain simulation, and reads the Q-analysis object. Post-processing determines the resonance to be used for scoring, evaluates the far field at that wavelength, and records the result in a CSV log. Invalid runs retain their parameter vector and failure flag. Representative structures are generated from fresh FDTD copies for final checks.

\subsection{Bayesian Optimization}

For each objective value, a full FDTD run is necessary. The search employs a two-objective BoTorch loop with noisy expected hypervolume improvement \cite{Balandat2020BoTorch,Daulton2021qNEHVI}. Each evaluated design stores its offsets, Q-related objective, divergence-related objective, and scalar reporting score.

Two single-task Gaussian-process models are fitted with input normalization and output standardization. The bounded search region is either centered on the origin-template offset or, in local refinement mode, on the current best logged candidate. Recent BO work has also explored learned or adaptive kernels for limited-budget black-box optimization \cite{Wilson2016DeepKernelLearning,Suwandi2025CAKE}; here we keep the surrogate model fixed so that the FDTD and device-level verification remain the focus.

Acquisition selects the next FDTD query, and the scalar score ranks saved structures for reporting and rechecking. Because the variables are relative offsets, the search remains close to the recovered source template.

\subsection{Reliability-Aware Objective}

Lumerical's Q-analysis object returns both the fitted $Q$ and an uncertainty-related estimate $dQ$. Unstable fits are discounted through
\begin{equation}
Q_{\mathrm{eff}} =
\frac{Q}{1+\eta(dQ/Q)+\xi\max(dQ/Q-1,0)^2},
\label{eq:q_eff}
\end{equation}
The reported runs use $\eta=0.8$ and $\xi=1.2$. A stable decay fit ensures that the $Q_{\mathrm{eff}}$ remains close to the raw fitted $Q$, while a large relative fitting uncertainty reduces the retained value.

The online score follows three design choices fixed before the repeated runs: stay close to the \SI{1310}{nm} band, favour narrow far-field lobes, and limit the influence of a single large fitted $Q$. The complete scoring expression is given in Appendix~\ref{app:score}. BO uses the Q-related and divergence-related terms for acquisition, while the scalar score is kept for plotting and candidate auditing.

\subsection{Candidate Retention Protocol}

The final filter is stricter than the online acquisition window:
\begin{equation}
Q_{\mathrm{eff}}\geq10^6,\quad
|\lambda-\lambda_0|\leq\SI{2}{nm},\quad
\theta_{\mathrm{div}}\leq\SI{0.845}{deg}.
\label{eq:usable_filter}
\end{equation}
Logged high-Q points remain visible even when they fail this filter. Device-level comparison uses retained structures regenerated from the recovered FDTD template.

\subsection{Repeated Runs and Verification}

The log stores candidate offsets, physical metrics, score, and best-so-far values for every FDTD evaluation. It describes the search path; final device claims use standalone reruns from the recovered template. BO trajectories show time-to-usable-candidate behaviour, repeated seeds test persistence around one origin template, and rechecked FSP copies provide the final performance table.

Two additional FDTD templates test starting-template sensitivity. They are prepared by fixed geometry offsets through the same \texttt{pcsel} model path used by the optimizer, and they keep the 80-evaluation budget, bounds, score, and usable-candidate filter from the origin-template campaign. This checks whether near-\SI{1310}{nm}, low-divergence, million-level $Q_{\mathrm{eff}}$ candidates can still be recovered after the initial source file is changed within the same design family.

\subsection{Control Experiments}

Control runs separate BO behavior from the quality of the chosen local parameter bounds. Latin-hypercube sampling provides an unguided space-filling baseline. Differential evolution provides an adaptive derivative-free control under the same FDTD budget and score. Two seeds are completed for each control method. A small BO hyperparameter check keeps the same budget, source file, search bounds, score, and strict filter, but changes the initial design from four evaluations to eight baseline-plus-random evaluations.

We report the best logged score, best retained quality, usable-hit count, and first usable evaluation. This separates one-off high values from repeated production of candidates that satisfy the joint filter.

%% file: sections/04_results_bo.tex
\section{Results and Discussion}
\label{sec:results}

\subsection{Experimental Design}

Unless noted otherwise, results use the recovered \SI{1310}{nm} PCSEL source model and the same FDTD post-processing pipeline. Optimization logs describe the search path. Final device numbers come from fresh simulations started from copied source files. A threshold-sensitivity audit is performed on nearby wavelength-window choices.

All variables are scripted offsets applied to the baseline PCSEL geometry, matching the practical case where a reasonable template is available and manual sweeps are expensive. Repeated BO seeds probe stochastic variation around the origin. Different-start tests and same-budget controls measure the sensitivity of the local parameter box from two directions.

\subsection{Origin-Template BO Repeats}

Before optimization, the recovered source model is evaluated with the same FDTD analysis pipeline. The baseline resonance is already near the target band, so the task is local refinement: improve the cavity response while preserving wavelength alignment and low-divergence surface emission.

The baseline used for final reporting has $Q_{\mathrm{eff}}=7.17\times10^4$ and $dQ/Q=0.88$ (\Cref{tab:final_comparison}).

We then ran three BO initializations with the same source model, geometry-update script, and 80-evaluation budget. Different starting structures are treated separately. The usable-hit count uses the strict filter in \Cref{eq:usable_filter}.
\begin{figure}[tbp]
  \centering
  \includegraphics[width=0.98\linewidth]{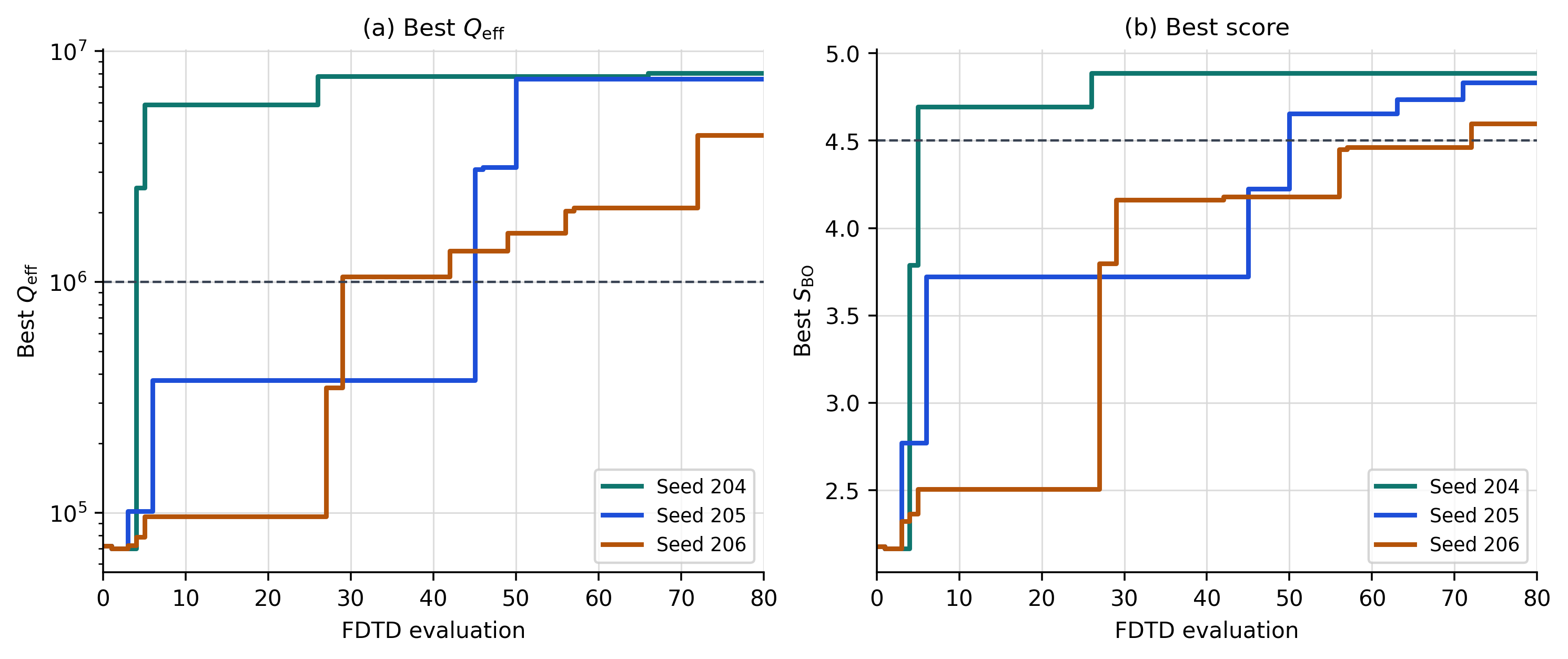}
  \caption{Repeated BO behavior under the same 80-evaluation budget. The panels track best-so-far $Q_{\mathrm{eff}}$ and best-so-far joint BO score from the common baseline.}
  \label{fig:bo_four_run_summary}
\end{figure}

\begin{table}[tbp]
  \centering
  \caption{Three BO repeated runs using the recovered source model and the same geometry-update script. ``Usable hits'' satisfy the joint filter in \Cref{eq:usable_filter}.}
  \label{tab:bo_four_run_summary}
  \resizebox{\linewidth}{!}{%
  \begin{tabular}{lcccccccc}
    \toprule
    Run & Invalid & Usable hits & First usable & Best eval. & Best $Q_{\mathrm{eff}}$ & $\lambda$ (nm) & $|\Delta\lambda|$ (nm) & Score \\
    \midrule
    Seed 204 & 2 & 15 & 5 & 26 & $7.77\times10^6$ & 1309.37 & 0.63 & 4.89 \\
    Seed 205 & 8 & 7 & 60 & 71 & $6.89\times10^6$ & 1310.90 & 0.90 & 4.83 \\
    Seed 206 & 4 & 5 & 29 & 72 & $4.33\times10^6$ & 1308.23 & 1.77 & 4.60 \\
    \bottomrule
  \end{tabular}
  }
\end{table}

The three repeats differ enough to report separately. Seed 204 reaches the usable region at evaluation 5 and leaves 15 candidates that pass the joint filter. Seed 206 reaches the filter at evaluation 29. Seed 205 is slower, with its first hit at evaluation 60, but the late route still ends with $Q_{\mathrm{eff}}=6.89\times10^6$. Across the three runs, BO repeatedly recovers near-\SI{1310}{nm}, sub-degree, million-level $Q_{\mathrm{eff}}$ candidates within an 80-run budget.

\subsection{Changed Starting Templates}

For starting-template sensitivity, two FDTD templates were created by applying finite offsets through the same geometry script. The altered starts stay within the local design family. Each run again uses an 80-evaluation budget and the joint filter in \Cref{eq:usable_filter}.

\begin{figure}[tbp]
  \centering
  \includegraphics[width=0.98\linewidth]{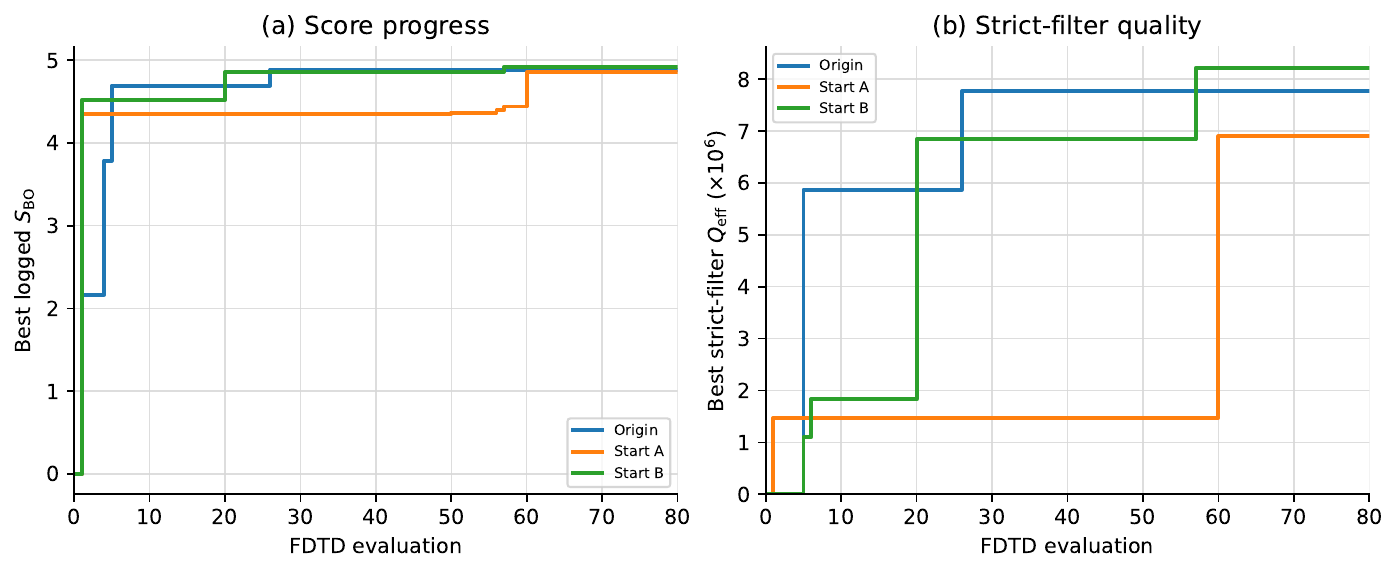}
  \caption{BO behavior from the original starting template and two prepared FDTD templates. All runs use the same geometry-update script, score, bounds, and candidate-retention rule.}
  \label{fig:bo_multistart_pcsel}
\end{figure}

\begin{table}[tbp]
  \centering
  \caption{Original-start reference and different-start BO runs under the same 80-evaluation budget and joint candidate filter.}
  \label{tab:bo_multistart_pcsel}
  \resizebox{\linewidth}{!}{%
  \begin{tabular}{lcccccccc}
    \toprule
    Start & Invalid & Usable hits & First usable & Best eval. & Best $Q_{\mathrm{eff}}$ & $\lambda$ (nm) & Divergence (deg) & Score \\
    \midrule
    Origin & 2 & 15 & 5 & 26 & $7.77\times10^6$ & 1309.37 & 0.8404 & 4.89 \\
    A & 9 & 4 & 1 & 60 & $6.90\times10^6$ & 1310.51 & 0.8420 & 4.85 \\
    B & 4 & 17 & 5 & 57 & $8.22\times10^6$ & 1310.13 & 0.8409 & 4.92 \\
    \bottomrule
  \end{tabular}
  }
\end{table}

Both prepared starts recover near-target, sub-degree, high-$Q_{\mathrm{eff}}$ candidates. Start A is the cautionary case: its best candidate is strong, but the run produces only four hits. Start B gives both the largest retained $Q_{\mathrm{eff}}$ and the highest usable-hit count in this group. The best retained quality is less start-dependent than the number of usable candidates produced along the way.

\subsection{Same-Budget Controls and Objective Selection}

Controls consist of two 80-evaluation Latin-hypercube runs and two 80-evaluation differential-evolution runs, both evaluated with the same score and joint filter. A BO variant with eight initial evaluations is included as a compact hyperparameter check. These runs test whether the local parameter box itself, rather than BO alone, contains good candidates.

\begin{figure}[tbp]
  \centering
  \includegraphics[width=0.98\linewidth]{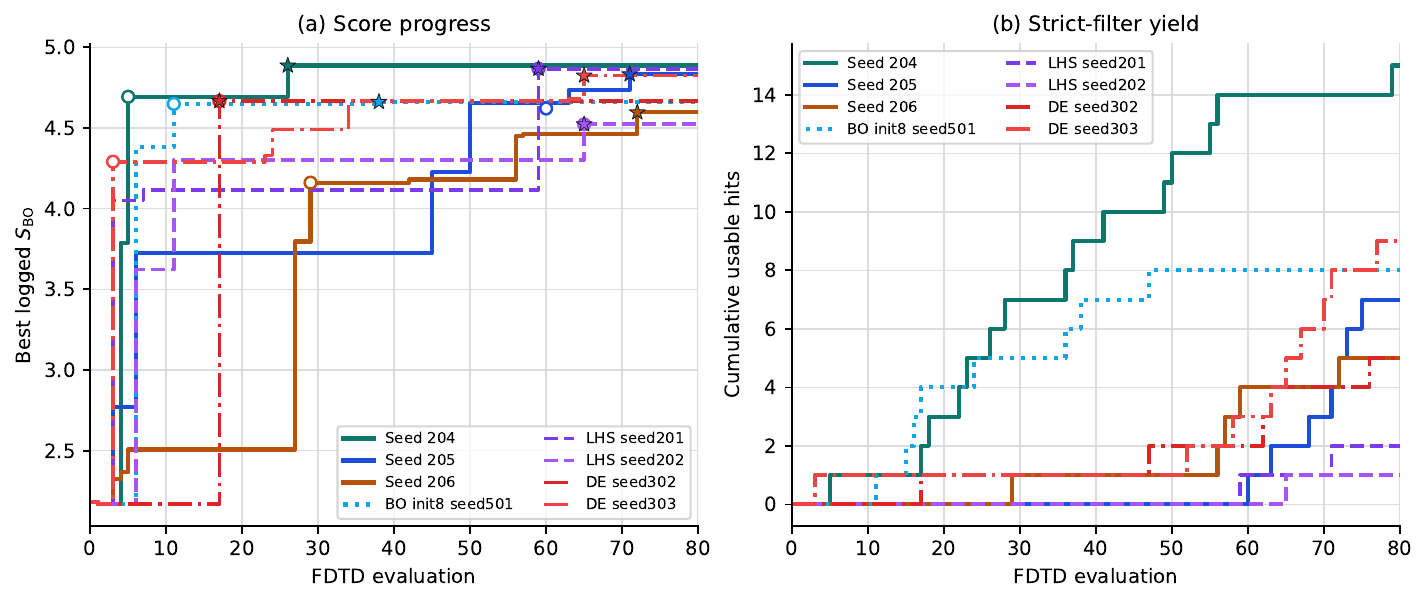}
  \caption{Same-budget BO and control comparison. The panels show best logged score and cumulative joint-filter hits; open circles mark first hits and stars mark best retained candidates.}
  \label{fig:same_budget_controls}
\end{figure}

\begin{table}[!t]
  \centering
  \caption{Compact same-budget comparison. Full per-run values are listed in Appendix~\ref{app:extra_results}.}
  \label{tab:same_budget_controls}
  \footnotesize
  \begin{tabular}{llccc}
    \toprule
    Run & Method & First usable & Hits & Best $Q_{\mathrm{eff}}$ \\
    \midrule
    Seed 204 & BO & 5 & 15 & $7.77\times10^6$ \\
    Seed 205 & BO & 60 & 7 & $6.89\times10^6$ \\
    Seed 206 & BO & 29 & 5 & $4.33\times10^6$ \\
    Init8 501 & BO & 11 & 8 & $4.62\times10^6$ \\
    LHS 201 & LHS & 59 & 2 & $7.00\times10^6$ \\
    LHS 202 & LHS & 65 & 1 & $2.90\times10^6$ \\
    DE 302 & DE & 17 & 5 & $3.95\times10^6$ \\
    DE 303 & DE & 3 & 9 & $7.69\times10^6$ \\
    \bottomrule
  \end{tabular}
\end{table}

One LHS run finds a late high-$Q_{\mathrm{eff}}$ candidate, and DE can reach a usable basin early. BO seed 204 has the largest usable-hit count in the main comparison, and the eight-initial-point BO variant reaches its first hit by evaluation 11. The control comparison should therefore be read through yield, not only through a single peak value. The stronger BO evidence is the number of geometries that survive the strict wavelength, divergence, and Q-fit filters.

A $Q_{\mathrm{eff}}$-only rule would select a different retained structure in six of the eight logs. In seeds 204 and 205, the largest logged $Q_{\mathrm{eff}}$ points sit just outside the \SI{2}{nm} wavelength window, moving the retained candidates to evaluations 26 and 71. In the BO init8 run, the largest-$Q_{\mathrm{eff}}$ point is detuned by \SI{5.19}{nm}; the candidate at evaluation 38 stays inside the strict window. LHS seed 202 and both DE runs show the same tradeoff.

The hit-count trend is stable for \SI{1}{nm}--\SI{2}{nm} wavelength windows with the same divergence and $Q_{\mathrm{eff}}$ limits. The three BO repeats give the highest mean hit count, and the BO init8 variant has a similar yield. DE reaches the first usable point earlier because seed 303 has a hit at evaluation 3, but its mean hit count is lower. LHS has the lowest yield despite one excellent late candidate.

\subsection{Parameter-Response Audit}

For the parameter audit, valid candidates are pooled from the three BO repeats, the BO init8 run, and the repeated LHS/DE controls. The data set contains 579 valid FDTD evaluations, including 52 candidates passing the joint filter. Each variable is normalized by its search bound, so the trends describe movement inside the local design box.

\begin{figure}[tbp]
  \centering
  \includegraphics[width=0.98\linewidth]{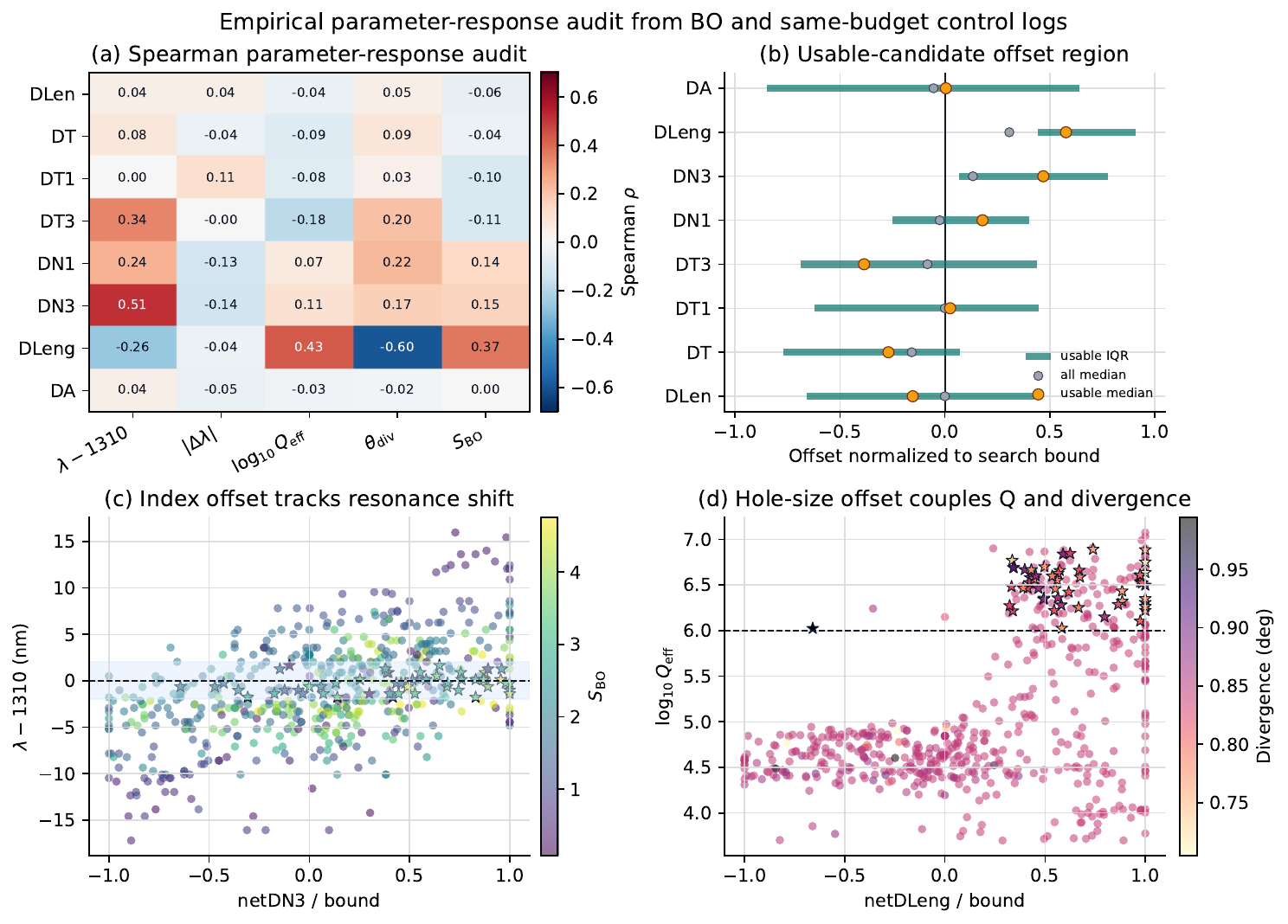}
  \caption{Empirical parameter-response audit using valid BO, BO-init8, LHS, and DE evaluations. The panels show rank correlations, usable-candidate offset ranges, and the two clearest response handles.}
  \label{fig:parameter_response_mechanism}
\end{figure}

The audit is a correlation analysis on the local design box. Its value is to show which handles successful searches keep using. \texttt{netDN3} is the wavelength handle: its correlation with $\lambda-1310$ is positive ($\rho=0.51$), as expected for an index-driven red shift. \texttt{netDLeng} is the more useful design handle. It raises $\log_{10}Q_{\mathrm{eff}}$ ($\rho=0.43$) while reducing far-field divergence ($\rho=-0.60$), which is the rare direction where the cavity and beam metrics improve together. The usable candidates therefore occupy positive \texttt{netDN1}, \texttt{netDN3}, and \texttt{netDLeng} offsets. Thickness and lattice offsets are less diagnostic in the pooled logs.

One metric is insufficient. Positive index offsets help place the resonance, but excessive red shift fails the wavelength filter. Hole-size offsets can reduce leakage and tighten the far-field lobe, while the robustness test shows that moving too far along this direction can switch the selected resonance and inflate $dQ/Q$.

\subsection{Independent FDTD Checks and Physical Fields}

Final device numbers come from standalone FDTD simulations. The seed-204/205/206 candidates were rebuilt from recovered source-model copies and processed with the same resonance, $Q$, $dQ/Q$, $Q_{\mathrm{eff}}$, and far-field pipeline. The early verified BO candidate raises $Q_{\mathrm{eff}}$ from $7.17\times10^4$ to $1.47\times10^6$ near \SI{1310}{nm}. The repeated-run candidates reach higher verified values in \Cref{tab:final_comparison}.

\begin{table}[tbp]
  \centering
  \caption{Representative regenerated FDTD checks evaluated from the recovered \texttt{PCSEL-1310-GaAs100\_origin.fsp} baseline.}
  \label{tab:final_comparison}
  \resizebox{\linewidth}{!}{%
  \begin{tabular}{llccccccc}
    \toprule
    Structure & Status & $\lambda$ (nm) & $Q$ & $dQ$ & $dQ/Q$ & Effective $Q$ & Divergence (deg) & BO score \\
    \midrule
    Origin baseline & baseline rerun & 1312.81 & $1.22\times10^5$ & $1.07\times10^5$ & 0.88 & $7.17\times10^4$ & 0.85 & 2.18 \\
    BO evaluation 2 & origin rerun & 1310.51 & $1.83\times10^6$ & $5.57\times10^5$ & 0.30 & $1.47\times10^6$ & 0.84 & 4.35 \\
    BO seed 204 eval. 26 & origin rerun & 1309.37 & $1.09\times10^7$ & $5.63\times10^6$ & 0.51 & $7.76\times10^6$ & 0.84 & 4.89 \\
    BO seed 205 eval. 71 & origin rerun & 1310.90 & $1.07\times10^7$ & $7.30\times10^6$ & 0.68 & $6.89\times10^6$ & 0.84 & 4.83 \\
    BO seed 206 eval. 72 & origin rerun & 1308.23 & $6.51\times10^6$ & $4.10\times10^6$ & 0.63 & $4.33\times10^6$ & 0.84 & 4.60 \\
    LHS seed 201 eval. 59 & control rerun & 1310.13 & $9.72\times10^6$ & $4.72\times10^6$ & 0.49 & $6.99\times10^6$ & 0.84 & 4.87 \\
    DE seed 302 eval. 17 & control rerun & 1309.37 & $8.05\times10^6$ & $9.85\times10^6$ & 1.22 & $3.95\times10^6$ & 0.84 & 4.67 \\
    \bottomrule
  \end{tabular}
  }
\end{table}

The regenerated FDTD runs are where the optimizer log either survives or fails. In \Cref{tab:final_comparison}, the LHS control survives well: it is wavelength centered and has high $Q_{\mathrm{eff}}$. The DE control remains competitive in $Q_{\mathrm{eff}}$, although its $dQ/Q=1.22$ is the warning sign. This failure mode is informative because it shows why the score discounts raw high-$Q$ fits before accepting a candidate. Candidate retention combines wavelength alignment, divergence, Q-fit reliability, and independent FDTD verification.

\begin{figure}[tbp]
  \centering
  \includegraphics[width=0.98\linewidth]{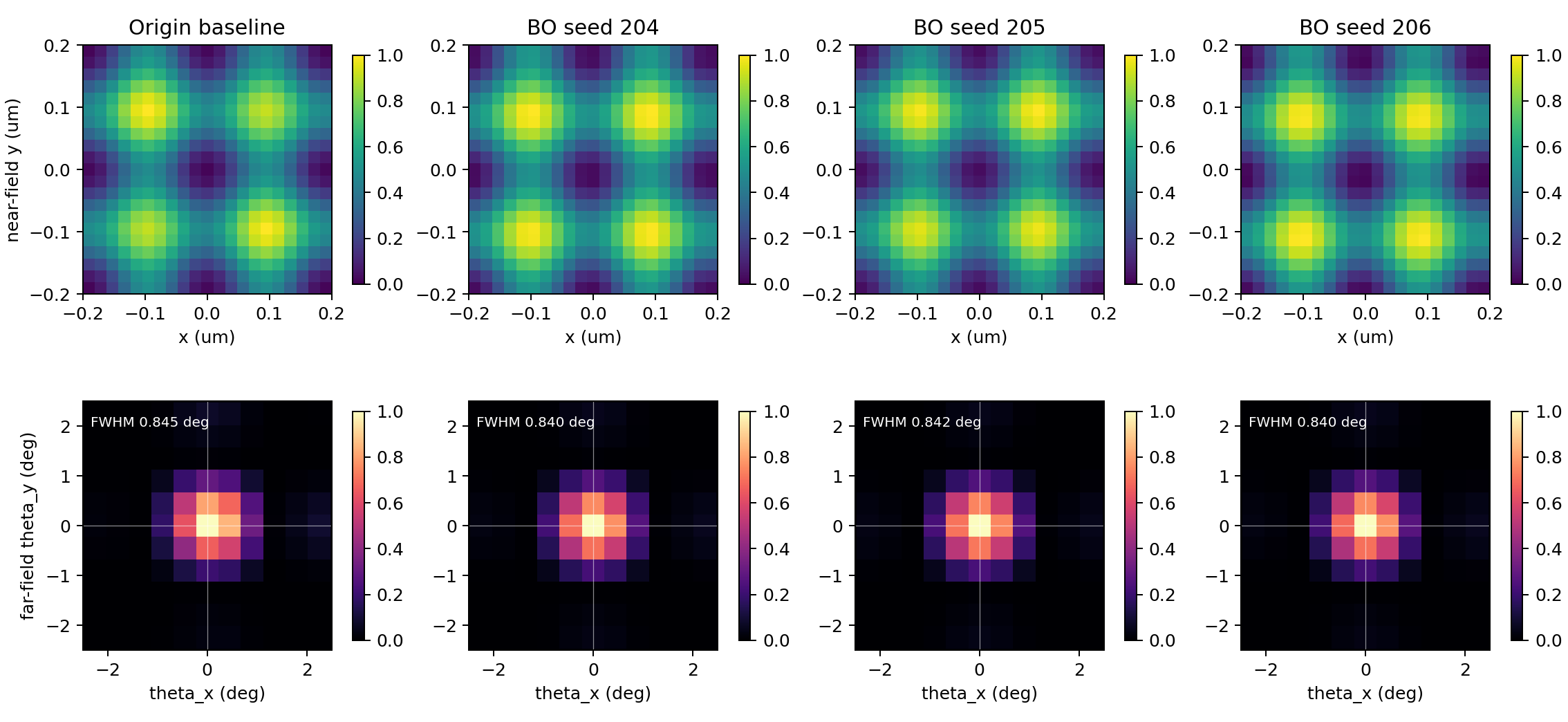}
  \caption{Field checks for the origin baseline and three verified BO repeated-run candidates, including near-field intensity and resonance-followed far-field intensity.}
  \label{fig:source_field_check}
\end{figure}

Field maps in \Cref{fig:source_field_check} compare the baseline with the three verified BO repeated-run candidates. The near-field panels keep the same mode family as the origin template. In the far field, the optimized candidates retain a surface-normal main lobe and slightly reduce the extracted FWHM divergence.

\begin{figure}[tbp]
  \centering
  \includegraphics[width=0.98\linewidth]{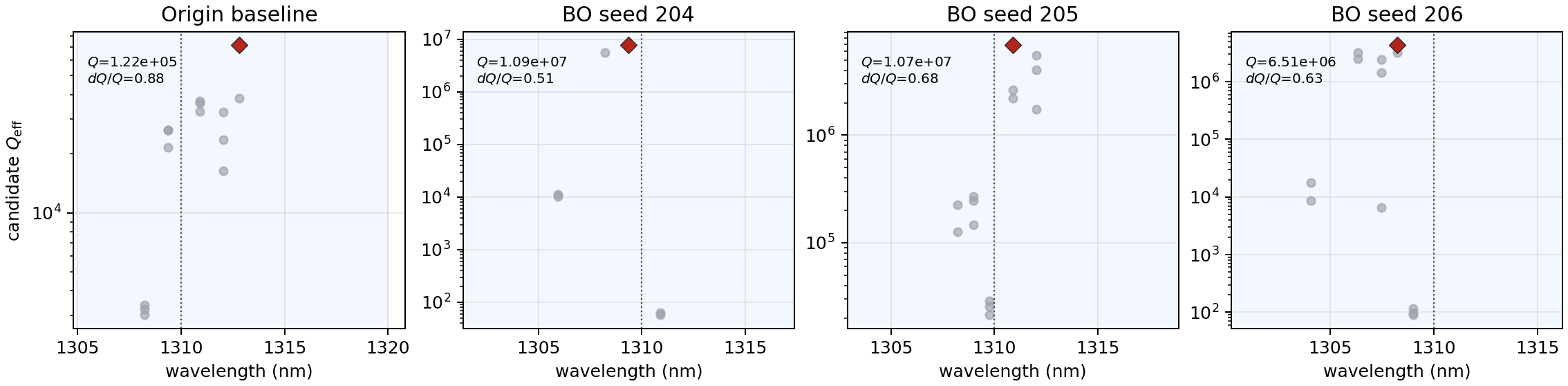}
  \caption{Q-analysis candidate maps for the saved source-model FSP files in \Cref{fig:source_field_check}. The red diamond marks the resonance selected for final reporting.}
  \label{fig:q_spectrum_comparison_bo}
\end{figure}

The Q-map panels in \Cref{fig:q_spectrum_comparison_bo} check the table values against nearby Q-analysis candidates. Final metrics combine $Q_{\mathrm{eff}}$ and $dQ/Q$ with the raw-$Q$ resonance map.

\subsection{Solver and Local Robustness Checks}

Seed 204 evaluation 26 was rerun under a small set of solver-setting changes. This representative check asks whether the candidate remains a near-target high-$Q$ mode when the FDTD time window or local mesh step is changed.

\begin{figure}[t]
  \centering
  \includegraphics[width=0.98\linewidth]{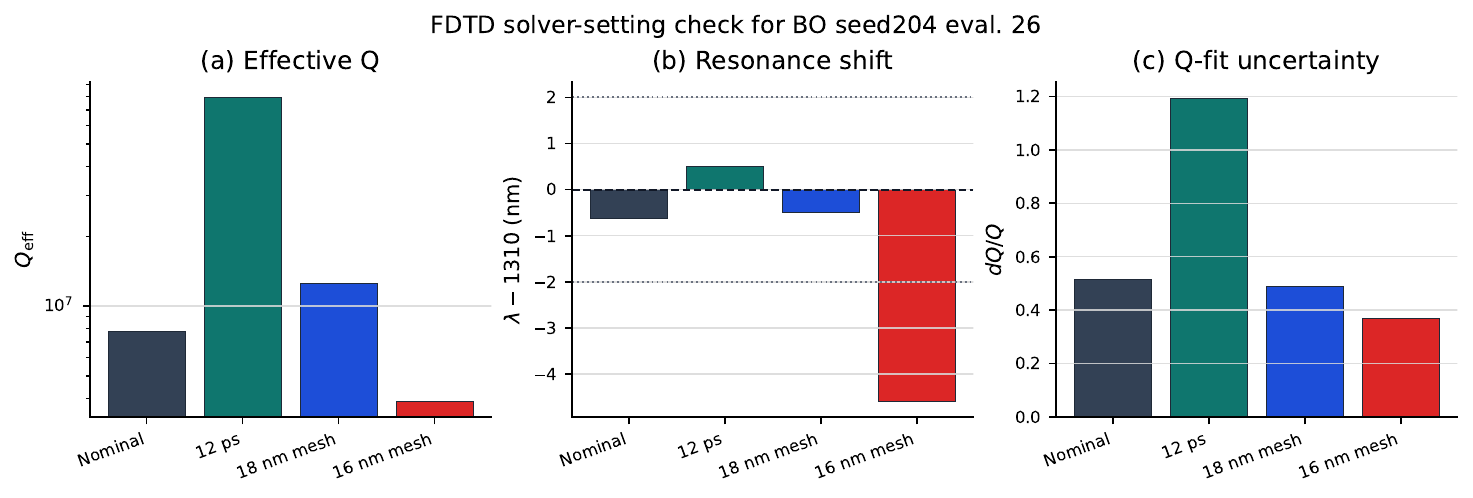}
  \caption{FDTD solver-setting check for BO seed 204 evaluation 26 under changes in simulation time and local mesh size.}
  \label{fig:seed204_convergence}
\end{figure}

In the solver-setting check, the nominal repeat reproduces the verification result. A longer time window and a moderate \SI{18}{nm} mesh refinement keep the selected resonance inside the wavelength window, although the longer run also increases $dQ/Q$. The stronger \SI{16}{nm} mesh changes the selected Q-analysis candidate to \SI{1305.40}{nm}. A fuller mode-tracked mesh convergence test is needed before device-ready claims.

\subsubsection*{Single-parameter robustness}

BO seed 204 evaluation 26 was further tested with small one-at-a-time perturbations to the scripted offsets. The edits cover positive and negative changes in the hole-size handle, lattice/period offset, thickness offset, and \texttt{netDN3} index offset.

\begin{figure}[t]
  \centering
  \includegraphics[width=0.98\linewidth]{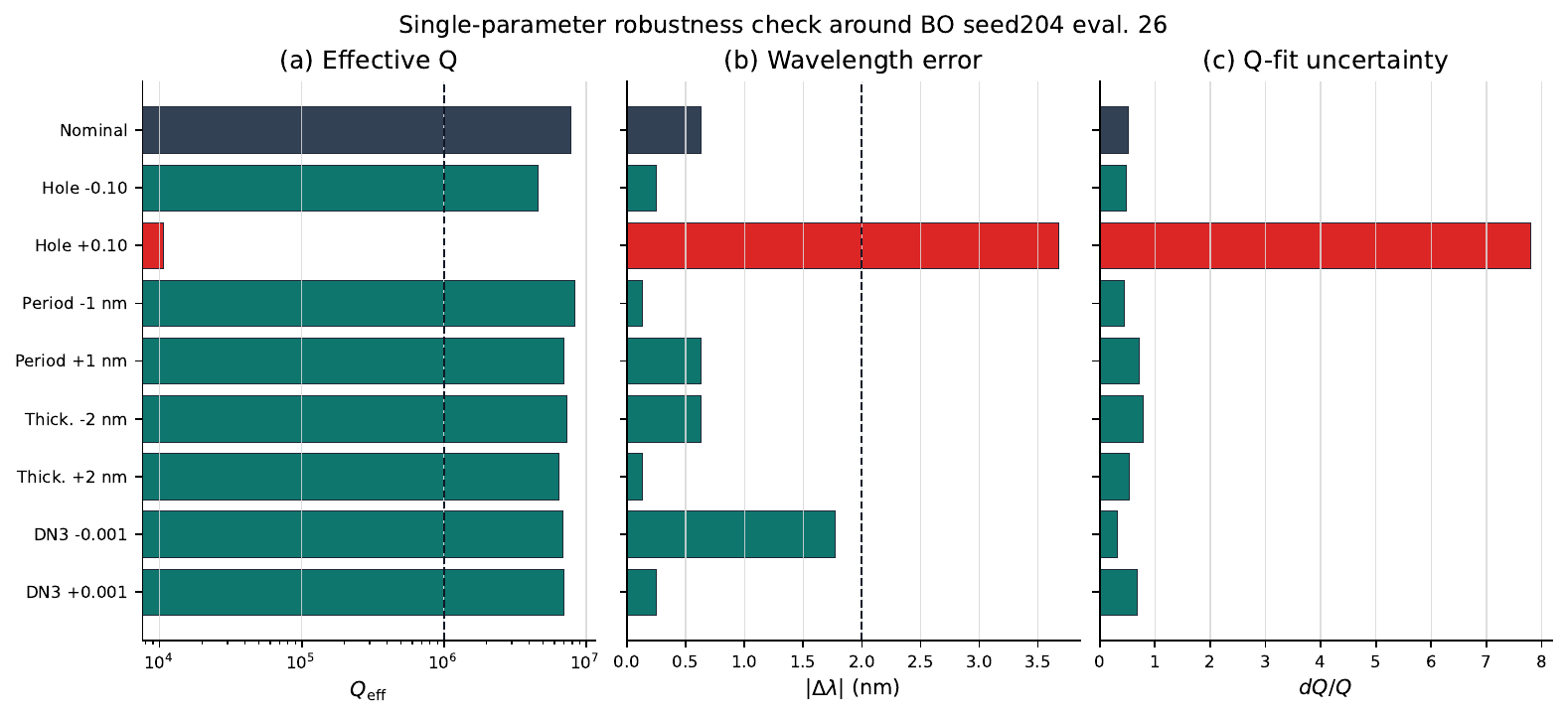}
  \caption{Single-parameter robustness check around BO seed 204 evaluation 26. Bar color marks pass or fail under the full strict usable filter.}
  \label{fig:seed204_robustness}
\end{figure}

Seven of the eight perturbed copies remain inside the strict filter (\Cref{fig:seed204_robustness}). Among the passing perturbations, the lowest $Q_{\mathrm{eff}}$ is $4.55\times10^6$, the largest wavelength error is \SI{1.77}{nm}, and the largest divergence is \SI{0.841}{deg}. The failed positive hole-size perturbation moves the selected resonance to \SI{1306.32}{nm}, increases $dQ/Q$ to 7.81, and reduces $Q_{\mathrm{eff}}$ to $1.06\times10^4$.

\subsection{Numerical Limits}

The parameter audit points to a narrow corridor, but there is no evidence of a monotonic optimum. Wavelength-compatible candidates cluster at positive \texttt{netDN1}, \texttt{netDN3}, and \texttt{netDLeng}. In the band-edge picture, \texttt{netDN3} shifts the resonance through the modal index, whereas \texttt{netDLeng} alters the air-fill factor and the dielectric Fourier component that governs vertical leakage. Moving in this direction helps until the resonance leaves the \SI{2}{nm} window or the Q-analysis routine locks onto a nearby poorly fitted peak.

The controls fit this reading. LHS seed 201 has a high-$Q_{\mathrm{eff}}$ candidate near \SI{1310}{nm}, and DE seed 303 reaches the usable region early. BO is separated less by a unique peak than by the number of saved geometries that pass the wavelength, divergence, and $dQ/Q$ checks in the same run. Later mesh, mode, and fabrication screens will discard additional structures, so a larger verified pool has practical value.

Mode identity is the numerical weak point. At the nominal settings, the optimized near fields stay in the origin-template mode family and the far field keeps a surface-normal lobe. At \SI{16}{nm} mesh, however, the selected Q-analysis point moves to \SI{1305.40}{nm}. The next verification should track seed 204 evaluation 26 through mesh and time-window changes using field-overlap continuity, then repeat the perturbation check with coupled changes in \texttt{netDLeng}, \texttt{netDA}, and \texttt{netDN3}. A small band-structure or coupled-wave calculation along that same path would turn the present correlation-based reading into a mode-level explanation.

\subsection{Physical Insight}

The field and spectrum checks show that the improvement is not merely a change in the optimizer score. The optimized near fields remain in the mode family observed in the reference model, and their far fields preserve a surface-normal main lobe. At the same time, verified $Q_{\mathrm{eff}}$ rises by roughly two orders of magnitude while the divergence changes only from about $0.85^{\circ}$ to $0.84^{\circ}$. The nearly fixed beam width is consistent with the unchanged lateral aperture: local unit-cell tuning primarily modifies the leakage rate and radiation distribution rather than the emitting area.

Two variables give the clearest device-level interpretation. The positive correlation between \texttt{netDN3} and wavelength ($\rho=0.51$) follows the increase in modal optical path produced by a larger index offset. The hole-size handle \texttt{netDLeng} correlates positively with $\log_{10}Q_{\mathrm{eff}}$ ($\rho=0.43$) and negatively with divergence ($\rho=-0.60$). Within this design box, the same change in air filling and dielectric modulation can therefore reduce vertical leakage while keeping the surface-emitted lobe narrow. These trends explain why the best region lies at positive \texttt{netDN3} and \texttt{netDLeng}, but they do not imply a universal monotonic rule.

The failed positive hole-size perturbation marks the boundary of that region. Its selected resonance moves from the target band to \SI{1306.32}{nm}, $dQ/Q$ rises to 7.81, and $Q_{\mathrm{eff}}$ collapses to $1.06\times10^4$. Together with the neighbouring peaks in the Q-spectrum map, this behaviour is consistent with a change in the resonance selected by the finite-time analysis. A practical tuning rule follows from the full-wave data: use the index-related offset to place the band-edge mode, use the hole-size offset to adjust leakage, and stop before the field pattern or decay fit loses continuity.

%% file: sections/05_conclusion_bo.tex
\section{Conclusion}
\label{sec:conclusion}

We used reliability-aware BO to refine eight scripted variables of a near-band \SI{1310}{nm} PCSEL model. Under the joint requirements $Q_{\mathrm{eff}}\geq10^6$, $|\lambda-\SI{1310}{nm}|\leq\SI{2}{nm}$, and $\theta_{\mathrm{div}}\leq\SI{0.845}{deg}$, each of the three 80-evaluation repeats produced 5--15 retained candidates. Independent FDTD reconstructions confirmed $Q_{\mathrm{eff}}=4.33\times10^6$--$7.76\times10^6$, corresponding to a 60--108-fold increase over the reference metric, with resonances at 1308.23--1310.90~nm and approximately $0.84^{\circ}$ divergence. The main outcome is a verified pool of high-$Q$ structures that satisfies wavelength, beam, and decay-fit conditions simultaneously.

The full-wave records also expose a useful local design rule. The index-related offset \texttt{netDN3} primarily places the band-edge resonance, whereas the hole-size-related \texttt{netDLeng} adjusts vertical leakage and the $Q$--divergence balance. Near-field continuity, a surface-normal far-field lobe, and seven passing single-parameter perturbations support this interpretation around the selected structure. The mode change observed for the \SI{16}{nm} mesh and the failed hole-size perturbation define the present numerical boundary: mode-tracked convergence remains necessary before fabrication-level claims. Within that boundary, the workflow turns a limited FDTD budget into physically checked PCSEL candidates.

%% file: sections/06_appendix.tex
\appendices

\newcommand{\appsinglefig}[4][\columnwidth]{%
  \par\smallskip
  \noindent\begin{minipage}{\columnwidth}
    \centering
    \includegraphics[width=#1]{#2}
    \refstepcounter{figure}\label{#4}
    \par\footnotesize Fig.~\thefigure. #3
  \end{minipage}
  \par\smallskip\normalsize
}

\section{Online Score}
\label{app:score}

The online score used during all BO, LHS, and DE runs is listed here for reproducibility. Let $\Delta\lambda=|\lambda-\lambda_0|$, $[u]_+=\max(u,0)$, and reject candidates outside the valid wavelength window $\Delta\lambda_{\mathrm{valid}}=\SI{20}{nm}$. For the remaining candidates,
\begin{equation}
\begin{aligned}
w_{\lambda} &=
\exp\left[-\frac{1}{2}
\left(\frac{\Delta\lambda}{\sigma_{\lambda}}\right)^2\right],
\quad \sigma_{\lambda}=\SI{8}{nm},\\
\widetilde{G}_Q &=
\left[\log_{10}
\frac{\min(Q_{\mathrm{eff}},Q_{\mathrm{cap}})}{Q_0}+1\right]_+^{1.25}\\
&\quad
+0.75\log_{10}\max\left(\frac{Q_{\mathrm{eff}}}{Q_{\mathrm{tar}}},1\right)
-0.06\left(\frac{\Delta\lambda}{\sigma_{\lambda}}\right)^2,\\
G_Q &= w_{\lambda}\left[\widetilde{G}_Q\right]_+,\\
G_{\theta} &=
w_{\lambda}\mathbb{I}_{\mathrm{div}}
\left[
0.65\frac{\theta_0}{\max(\theta_{\mathrm{div}},\theta_0)}
+0.20B_{\theta}\right.\\
&\quad\left.
+0.15\min\left(\frac{C}{C_0},1.5\right)
\right],\\
S_{\mathrm{BO}} &=
G_Q+0.55G_{\theta}
+0.15w_{\lambda}\mathbb{I}_Q
+0.10\mathbb{I}_Q\mathbb{I}_{\theta}.
\end{aligned}
\label{eq:bo_score}
\end{equation}
The constants are $Q_{\mathrm{cap}}=3\times10^5$, $Q_0=2\times10^4$, $Q_{\mathrm{tar}}=10^5$, $\theta_0=\SI{6}{deg}$, and $C_0=0.30$. The center-cone power fraction is $C$. The indicators are $\mathbb{I}_Q=1$ for $Q_{\mathrm{eff}}\geq Q_{\mathrm{tar}}$, $\mathbb{I}_{\theta}=1$ for $\theta_{\mathrm{div}}\leq\SI{12}{deg}$, and $\mathbb{I}_{\mathrm{div}}=1$ for $Q_{\mathrm{eff}}\geq1.5\times10^4$. The divergence soft term is $B_{\theta}=1$ for $\theta_{\mathrm{div}}\leq\SI{12}{deg}$ and $B_{\theta}=12/\theta_{\mathrm{div}}$ otherwise.

\section{Same-Budget Details}
\label{app:extra_results}

\par\smallskip
\noindent\begin{minipage}{\columnwidth}
  \refstepcounter{table}\label{tab:same_budget_controls_full_a}
  \centering
  {\footnotesize TABLE~\thetable\par}
  {\scriptsize SAME-BUDGET COMPARISON: SCORE, TIMING, AND YIELD.\par}
  \vspace{2pt}
  \scriptsize
  \setlength{\tabcolsep}{2.6pt}
  \begin{tabular}{lccccc}
    \toprule
    Run & Method & Eval. & Score & Hits & First \\
    \midrule
    Seed 204 & BO & 26 & 4.89 & 15 & 5 \\
    Seed 205 & BO & 71 & 4.83 & 7 & 60 \\
    Seed 206 & BO & 72 & 4.60 & 5 & 29 \\
    Init8 501 & BO & 38 & 4.66 & 8 & 11 \\
    LHS 201 & LHS & 59 & 4.87 & 2 & 59 \\
    LHS 202 & LHS & 65 & 4.52 & 1 & 65 \\
    DE 302 & DE & 17 & 4.67 & 5 & 17 \\
    DE 303 & DE & 65 & 4.82 & 9 & 3 \\
    \bottomrule
  \end{tabular}
\end{minipage}
\par\smallskip

\par\smallskip
\noindent\begin{minipage}{\columnwidth}
  \refstepcounter{table}\label{tab:same_budget_controls_full_b}
  \centering
  {\footnotesize TABLE~\thetable\par}
  {\scriptsize SAME-BUDGET COMPARISON: RETAINED PHYSICAL METRICS.\par}
  \vspace{2pt}
  \scriptsize
  \setlength{\tabcolsep}{2.8pt}
  \begin{tabular}{lccc}
    \toprule
    Run & $Q_{\mathrm{eff}}$ & $\lambda$ (nm) & Div. (deg) \\
    \midrule
    Seed 204 & $7.77\times10^6$ & 1309.37 & 0.8404 \\
    Seed 205 & $6.89\times10^6$ & 1310.90 & 0.8419 \\
    Seed 206 & $4.33\times10^6$ & 1308.23 & 0.8397 \\
    Init8 501 & $4.62\times10^6$ & 1308.61 & 0.8406 \\
    LHS 201 & $7.00\times10^6$ & 1310.13 & 0.8413 \\
    LHS 202 & $2.90\times10^6$ & 1311.28 & 0.8417 \\
    DE 302 & $3.95\times10^6$ & 1309.37 & 0.8407 \\
    DE 303 & $7.69\times10^6$ & 1308.61 & 0.8404 \\
    \bottomrule
  \end{tabular}
\end{minipage}
\par\smallskip

\section{Supplementary Search Figures}
\label{app:supplementary_figures}

\appsinglefig[0.92\columnwidth]{appendix/bo_vs_baseline_best_qeff.pdf}
{Supplementary BO improvement against the baseline.}
{fig:app_bo_vs_baseline}

\appsinglefig[0.92\columnwidth]{appendix/raw_vs_effective_q.pdf}
{Supplementary comparison between raw fitted $Q$ and $Q_{\mathrm{eff}}$.}
{fig:app_raw_vs_effective}

\appsinglefig[0.92\columnwidth]{appendix/pcsel_candidate_landscape.pdf}
{Supplementary candidate landscape across repeated BO runs.}
{fig:app_candidate_landscape}

\appsinglefig[0.88\columnwidth]{appendix/pcsel_seed204_main_trajectory.pdf}
{Supplementary seed-204 BO trajectory.}
{fig:app_seed204_trajectory}

\appsinglefig[0.88\columnwidth]{appendix/pcsel_multiseed_hit_statistics.pdf}
{Supplementary hit statistics for repeated BO runs.}
{fig:app_hit_statistics}

\appsinglefig[0.90\columnwidth]{appendix/pcsel_objective_selection_audit.pdf}
{Supplementary objective-selection audit.}
{fig:app_objective_audit}

\appsinglefig[0.88\columnwidth]{appendix/threshold_sensitivity_summary.pdf}
{Supplementary sensitivity to wavelength-window choices.}
{fig:app_threshold_sensitivity}

\appsinglefig[0.88\columnwidth]{appendix/verified_qeff_dq_ratio.pdf}
{Supplementary verified $Q_{\mathrm{eff}}$ and $dQ/Q$ summary.}
{fig:app_verified_qeff}

%% file: sections/07_declarations.tex
\section*{Data Availability}

The FDTD parameter logs, post-processing scripts, and manuscript plotting data are available from the corresponding authors upon reasonable request. Commercial solver source files are subject to software-license and project-sharing restrictions.

\section*{Acknowledgment}

OpenAI Codex was used for language editing and revision suggestions throughout the manuscript and to assist with experimental scripting, data-processing checks, and modifications to figure and analysis code. The authors independently verified and finalized all scientific content, code changes, simulation settings, reported results, figures, references, and conclusions. No data, figures, references, or experimental results were fabricated by generative AI.

%% file: references.bib
@article{Sakurai2019ProgressPCSEL,
  author = {Sakurai, Akihiko and Ishizaki, Kenji and De Zoysa, Menaka and Tanaka, Yoshinori and Noda, Susumu},
  title = {Progress in Photonic-Crystal Surface-Emitting Lasers},
  journal = {Photonics},
  volume = {6},
  number = {3},
  pages = {96},
  year = {2019},
  doi = {10.3390/photonics6030096}
}

@article{Noda2023TutorialPCSEL,
  author = {Noda, Susumu and others},
  title = {High-power and high-beam-quality photonic-crystal surface-emitting lasers: a tutorial},
  journal = {Advances in Optics and Photonics},
  volume = {15},
  number = {4},
  pages = {977--1032},
  year = {2023},
  doi = {10.1364/AOP.490118}
}

@article{Hirose2014WattClassPCSEL,
  author = {Hirose, Kazuyoshi and Liang, Yong and Kurosaka, Yoshitaka and Watanabe, Akiyoshi and Sugiyama, Takahiro and Noda, Susumu},
  title = {Watt-class high-power, high-beam-quality photonic-crystal lasers},
  journal = {Nature Photonics},
  volume = {8},
  pages = {406--411},
  year = {2014},
  doi = {10.1038/nphoton.2014.75}
}

@article{Yoshida2019DoubleLatticePCSEL,
  author = {Yoshida, Masahiro and De Zoysa, Menaka and Ishizaki, Kenji and Tanaka, Yoshinori and Kawasaki, Masato and Hatsuda, Ranko and Song, Bongshik and Gelleta, John and Noda, Susumu},
  title = {Double-lattice photonic-crystal resonators enabling high-brightness semiconductor lasers with symmetric narrow-divergence beams},
  journal = {Nature Materials},
  volume = {18},
  pages = {121--128},
  year = {2019},
  doi = {10.1038/s41563-018-0242-y}
}

@article{Orchard2023SmallSignalPCSEL,
  author = {Orchard, Jonathan R. and Ivanov, Pavlo and McKenzie, Adam F. and Hill, Calum H. and Javed, Ibrahim and Munro, Connor W. and Kettle, Jeff and Hogg, Richard A. and Childs, David T. D. and Taylor, Richard J. E.},
  title = {Small signal modulation of photonic crystal surface emitting lasers},
  journal = {Scientific Reports},
  volume = {13},
  pages = {19019},
  year = {2023},
  doi = {10.1038/s41598-023-45414-7}
}

@article{Molesky2018InverseDesign,
  author = {Molesky, Sean and Lin, Zin and Piggott, Alexander Y. and Jin, Weiliang and Vuckovic, Jelena and Rodriguez, Alejandro W.},
  title = {Inverse design in nanophotonics},
  journal = {Nature Photonics},
  volume = {12},
  pages = {659--670},
  year = {2018},
  doi = {10.1038/s41566-018-0246-9}
}

@article{Liu2021MLPhotonicInverseDesign,
  author = {Liu, Dianjing and Tan, Yu and Khoram, Ehsan and Yu, Zongfu},
  title = {Training deep neural networks for the inverse design of nanophotonic structures},
  journal = {ACS Photonics},
  volume = {5},
  number = {4},
  pages = {1365--1369},
  year = {2018},
  doi = {10.1021/acsphotonics.7b01377}
}

@article{Pan2023DeepAdjointReview,
  author = {Pan, Zongyong and Pan, Xiaomin},
  title = {Deep Learning and Adjoint Method Accelerated Inverse Design in Photonics: A Review},
  journal = {Photonics},
  volume = {10},
  number = {7},
  pages = {852},
  year = {2023},
  doi = {10.3390/photonics10070852}
}

@article{Minkov2014AutomatedPhCCavities,
  author = {Minkov, Momchil and Savona, Vincenzo},
  title = {Automated Optimization of Photonic Crystal Slab Cavities},
  journal = {Scientific Reports},
  volume = {4},
  pages = {5124},
  year = {2014},
  doi = {10.1038/srep05124}
}

@article{Shahriari2016BayesianOptimization,
  author = {Shahriari, Bobak and Swersky, Kevin and Wang, Ziyu and Adams, Ryan P. and de Freitas, Nando},
  title = {Taking the human out of the loop: A review of Bayesian optimization},
  journal = {Proceedings of the IEEE},
  volume = {104},
  number = {1},
  pages = {148--175},
  year = {2016},
  doi = {10.1109/JPROC.2015.2494218}
}

@inproceedings{Balandat2020BoTorch,
  author = {Balandat, Maximilian and Karrer, Brian and Jiang, Daniel R. and Daulton, Samuel and Letham, Benjamin and Wilson, Andrew Gordon and Bakshy, Eytan},
  title = {{BoTorch}: A Framework for Efficient Monte-Carlo Bayesian Optimization},
  booktitle = {Advances in Neural Information Processing Systems},
  volume = {33},
  pages = {21524--21538},
  year = {2020}
}

@inproceedings{Daulton2021qNEHVI,
  author = {Daulton, Samuel and Balandat, Maximilian and Bakshy, Eytan},
  title = {Parallel Bayesian Optimization of Multiple Noisy Objectives with Expected Hypervolume Improvement},
  booktitle = {Advances in Neural Information Processing Systems},
  volume = {34},
  pages = {2187--2200},
  year = {2021}
}

@inproceedings{Suwandi2025CAKE,
  author = {Suwandi, Richard and Yin, Feng and Wang, Juntao and Li, Renjie and Chang, Tsung-Hui and Theodoridis, Sergios},
  title = {Adaptive Kernel Design for Bayesian Optimization Is a Piece of {CAKE} with {LLM}s},
  booktitle = {Advances in Neural Information Processing Systems},
  volume = {38},
  pages = {132690--132723},
  year = {2025},
  url = {https://proceedings.neurips.cc/paper_files/paper/2025/file/c03a2610bca2712b984b331fd4f7bb6f-Paper-Conference.pdf}
}

@misc{AnsysQFactorCavity,
  author = {{Ansys Optics}},
  title = {Quality Factor Calculations for a Resonant Cavity},
  year = {2025},
  howpublished = {\url{https://optics.ansys.com/hc/en-us/articles/360041611774-Quality-factor-calculations-for-a-resonant-cavity}},
  note = {Accessed 2026-05-28}
}

@article{Imada1999CoherentPCSEL,
  author = {Imada, Masahiro and Noda, Susumu and Chutinan, Alongkarn and Tokuda, Toshihiko and Murata, Masahiko and Sasaki, Gosei},
  title = {Coherent two-dimensional lasing action in surface-emitting laser with triangular-lattice photonic crystal structure},
  journal = {Applied Physics Letters},
  volume = {75},
  number = {3},
  pages = {316--318},
  year = {1999},
  doi = {10.1063/1.124361}
}

@article{Noda2001PolarizationPCSEL,
  author = {Noda, Susumu and Yokoyama, Masayuki and Imada, Masahiro and Chutinan, Alongkarn and Mochizuki, Makoto},
  title = {Polarization Mode Control of Two-Dimensional Photonic Crystal Laser by Unit Cell Structure Design},
  journal = {Science},
  volume = {293},
  number = {5532},
  pages = {1123--1125},
  year = {2001},
  doi = {10.1126/science.1061738}
}

@article{Matsubara2008GaNPCSEL,
  author = {Matsubara, Hideki and Yoshimoto, Shin and Saito, Hiroyuki and Jianglin, Xie and Asano, Takashi and Noda, Susumu},
  title = {{GaN} Photonic-Crystal Surface-Emitting Laser at Blue-Violet Wavelengths},
  journal = {Science},
  volume = {319},
  number = {5862},
  pages = {445--447},
  year = {2008},
  doi = {10.1126/science.1150413}
}

@article{Noda2017JSTQEPCSEL,
  author = {Noda, Susumu and Kitamura, Kenji and Okino, Takashi and Yasuda, Daisuke and Tanaka, Yoshinori},
  title = {Photonic-Crystal Surface-Emitting Lasers: Review and Introduction of Modulated-Photonic Crystals},
  journal = {IEEE Journal of Selected Topics in Quantum Electronics},
  volume = {23},
  number = {6},
  pages = {4900107},
  year = {2017},
  doi = {10.1109/JSTQE.2017.2696883}
}

@article{Inoue2020EnhancedFeedbackPCSEL,
  author = {Inoue, Takuya and Yoshida, Masahiro and De Zoysa, Menaka and Ishizaki, Kenji and Noda, Susumu},
  title = {Design of photonic-crystal surface-emitting lasers with enhanced in-plane optical feedback for high-speed operation},
  journal = {Optics Express},
  volume = {28},
  number = {4},
  pages = {5050--5057},
  year = {2020},
  doi = {10.1364/OE.385277}
}

@article{Itoh2020InPPCSEL,
  author = {Itoh, Yoshiaki and Yoshida, Masahiro and De Zoysa, Menaka and Ishizaki, Kenji and Noda, Susumu},
  title = {Continuous-wave lasing operation of 1.3-$\mu$m wavelength {InP}-based photonic crystal surface-emitting lasers using {MOVPE} regrowth},
  journal = {Optics Express},
  volume = {28},
  number = {24},
  pages = {35483--35491},
  year = {2020},
  doi = {10.1364/OE.404605}
}

@article{Sakata2020DuallyModulatedPCSEL,
  author = {Sakata, Ryo and Ishizaki, Kenji and De Zoysa, Menaka and Kitamura, Kenji and Inoue, Takuya and Gelleta, John and Noda, Susumu},
  title = {Dually modulated photonic crystals enabling high-power high-beam-quality two-dimensional beam scanning lasers},
  journal = {Nature Communications},
  volume = {11},
  pages = {3487},
  year = {2020},
  doi = {10.1038/s41467-020-17092-w}
}

@article{Yoshida2021LiDARPCSEL,
  author = {Yoshida, Masahiro and De Zoysa, Menaka and Ishizaki, Kenji and Noda, Susumu},
  title = {Photonic-crystal lasers with high-quality narrow-divergence symmetric beams and their application to {LiDAR}},
  journal = {Journal of Physics: Photonics},
  volume = {3},
  number = {2},
  pages = {022006},
  year = {2021},
  doi = {10.1088/2515-7647/abea06}
}

@article{Morita2021GainLossPCSEL,
  author = {Morita, Ryosuke and De Zoysa, Menaka and Ishizaki, Kenji and Noda, Susumu},
  title = {Photonic-crystal lasers with two-dimensionally arranged gain and loss sections for high-peak-power short-pulse operation},
  journal = {Nature Photonics},
  volume = {15},
  pages = {311--318},
  year = {2021},
  doi = {10.1038/s41566-021-00771-5}
}

@article{Inoue2022GeneralRecipePCSEL,
  author = {Inoue, Takuya and Yoshida, Masahiro and Gelleta, John and Izumi, Koki and Yoshida, Keisuke and Ishizaki, Kenji and De Zoysa, Menaka and Noda, Susumu},
  title = {General recipe to realize photonic-crystal surface-emitting lasers with 100-{W}-to-1-{kW} single-mode operation},
  journal = {Nature Communications},
  volume = {13},
  pages = {3262},
  year = {2022},
  doi = {10.1038/s41467-022-30910-7}
}

@article{Jensen2011TopologyNanophotonics,
  author = {Jensen, Jakob S{\o}ndergaard and Sigmund, Ole},
  title = {Topology optimization for nano-photonics},
  journal = {Laser \& Photonics Reviews},
  volume = {5},
  number = {2},
  pages = {308--321},
  year = {2011},
  doi = {10.1002/lpor.201000014}
}

@article{LalauKeraly2013AdjointShape,
  author = {Lalau-Keraly, Christopher M. and Bhargava, Samarth and Miller, Owen D. and Yablonovitch, Eli},
  title = {Adjoint shape optimization applied to electromagnetic design},
  journal = {Optics Express},
  volume = {21},
  number = {18},
  pages = {21693--21701},
  year = {2013},
  doi = {10.1364/OE.21.021693}
}

@article{Piggott2015WDMInverseDesign,
  author = {Piggott, Alexander Y. and Lu, Jesse and Lagoudakis, Konstantinos G. and Petykiewicz, Jan and Babinec, Thomas M. and Vu{\v{c}}kovi{\'c}, Jelena},
  title = {Inverse design and demonstration of a compact and broadband on-chip wavelength demultiplexer},
  journal = {Nature Photonics},
  volume = {9},
  number = {6},
  pages = {374--377},
  year = {2015},
  doi = {10.1038/nphoton.2015.69}
}

@article{Christiansen2021TopOptTutorial,
  author = {Christiansen, Rasmus E. and Sigmund, Ole},
  title = {Inverse design in photonics by topology optimization: tutorial},
  journal = {Journal of the Optical Society of America B},
  volume = {38},
  number = {2},
  pages = {496--509},
  year = {2021},
  doi = {10.1364/JOSAB.406048}
}

@article{Ma2021DeepLearningPhotonics,
  author = {Ma, Wei and Liu, Zhaocheng and Kudyshev, Zhaxylyk A. and Boltasseva, Alexandra and Cai, Wenshan and Liu, Yongmin},
  title = {Deep learning for the design of photonic structures},
  journal = {Nature Photonics},
  volume = {15},
  number = {2},
  pages = {77--90},
  year = {2021},
  doi = {10.1038/s41566-020-0685-y}
}

@article{Asano2019NanocavityDNN,
  author = {Asano, Takashi and Noda, Susumu},
  title = {Iterative optimization of photonic crystal nanocavity designs by using deep neural networks},
  journal = {Nanophotonics},
  volume = {8},
  number = {12},
  pages = {2243--2256},
  year = {2019},
  doi = {10.1515/nanoph-2019-0308}
}

@article{Sanchez2024MLOptimizationPhotonics,
  author = {Sanchez, M. and Everly, C. and Postigo, P. A.},
  title = {Advances in machine learning optimization for classical and quantum photonics},
  journal = {Journal of the Optical Society of America B},
  volume = {41},
  number = {2},
  pages = {A177--A190},
  year = {2024},
  doi = {10.1364/JOSAB.507268}
}

@article{Jones1998EGO,
  author = {Jones, Donald R. and Schonlau, Matthias and Welch, William J.},
  title = {Efficient Global Optimization of Expensive Black-Box Functions},
  journal = {Journal of Global Optimization},
  volume = {13},
  number = {4},
  pages = {455--492},
  year = {1998},
  doi = {10.1023/A:1008306431147}
}

@inproceedings{Snoek2012PracticalBO,
  author = {Snoek, Jasper and Larochelle, Hugo and Adams, Ryan P.},
  title = {Practical Bayesian Optimization of Machine Learning Algorithms},
  booktitle = {Advances in Neural Information Processing Systems},
  volume = {25},
  year = {2012}
}

@inproceedings{Wilson2016DeepKernelLearning,
  author = {Wilson, Andrew Gordon and Hu, Zhiting and Salakhutdinov, Ruslan and Xing, Eric P.},
  title = {Deep Kernel Learning},
  booktitle = {Proceedings of the 19th International Conference on Artificial Intelligence and Statistics},
  pages = {370--378},
  year = {2016}
}
